\documentstyle[12pt,epsf]{article} 
\begin{document}
\newcommand{\onefigure}[2]{\begin{figure}[htbp] 
         \begin{center}\leavevmode\epsfbox{#1.eps}\end{center}\caption{#2\label{#1}}
         \end{figure}}
\setlength{\parskip}{2ex}
\setlength{\parindent}{0em}
\setlength{\baselineskip}{2.5ex}
\newcommand{\be}{\begin{equation}}
\newcommand{\ee}{\end{equation}}
\newcommand{\bea}{\begin{eqnarray}}
\newcommand{\eea}{\end{eqnarray}}
\newcommand{\nn}{\nonumber}
\newcommand{\bean}{\begin{eqnarray*}}
\newcommand{\eean}{\end{eqnarray*}}
\newcommand{\half}{\frac{1}{2}}
\newcommand{\integ}[2]{\int\limits_{#1}^{#2}\!\!}
\newcommand{\del}{\partial}
\renewcommand{\Im}{\mbox{Im}}
\renewcommand{\arg}{\mbox{Arg}}
\newcommand{\dslash}{\not\partial}
\newcommand{\dm}{\partial_{\mu}}
\newcommand{\dn}{\partial_{\nu}}
\newcommand{\ds}{\partial_{\sigma}}
\newcommand{\dr}{\partial_{\rho}}
\newcommand{\gm}{\gamma_{\mu}}
\newcommand{\gn}{\gamma_{\nu}}
\newcommand{\gs}{\gamma_{\sigma}}
\newcommand{\gr}{\gamma_{\rho}}
\newcommand{\gum}{\gamma^{\mu}}
\newcommand{\gun}{\gamma^{\nu}}
\newcommand{\gus}{\gamma^{\sigma}}
\newcommand{\gur}{\gamma^{\rho}}
\newcommand{\e}{\varepsilon}
\newcommand{\gmn}{g^{\mu\nu}}
\newcommand{\gmnd}{g_{\mu\nu}}
\newcommand{\expect}[1]{\langle #1 \rangle}
\newcommand{\dd}{\mbox{d}}
\newcommand{\Gam}[1]{\Gamma\left(#1\right)}
\newcommand{\gE}{\gamma_E}
\newcommand{\ordo}[1]{{\cal O}(#1)}
\newcommand{\myref}[1]{(\ref{#1})}
\newcommand{\figref}[1]{Fig.~\protect\ref{#1}}
\newcommand{\secref}[1]{sec.~\protect\ref{#1}}
\newcommand{\idk}{\int\frac{\dd^dk}{(2\pi)^d}}
\newcommand{\dhalf}{\frac{d}{2}}
\newcommand{\fourpi}{(4\pi)^{\frac{d}{2}}}
\newcommand{\la}{\lambda^a}
\newcommand{\lb}{\lambda^b}
\newcommand{\lc}{\lambda^c}
\newcommand{\ld}{\lambda^d}
\newcommand{\fabc}{f^{abc}}
\newcommand{\dabc}{d^{abc}}
\newcommand{\unit}{1}
\newcommand{\mat}[4]{\left(\begin{array}{cc} #1 & #2 \\ #3 & #4 \end{array}\right)}
\newcommand{\vect}[2]{({#1\atop #2})}
\thispagestyle{empty}
\def\thefootnote{\fnsymbol{footnote}}
\begin{flushright}
  MIT-CTP-2740 \\
  hep-th/9805076
\end{flushright} \vskip 0.5cm

\begin{center}\LARGE
{\bf Equivalent String Networks and Uniqueness of BPS States}
\end{center} \vskip 0.8cm
\begin{center}\large
        Tam\'as Hauer 
\end{center}
\vskip0.3cm
\begin{center}
Center for Theoretical Physics\\
LNS and Department of Physics, MIT\\
Cambridge, MA 02139, USA
\end{center}
\begin{center}
E-mail: {\tt hauer@mit.edu}
\end{center}
\begin{center}
May 1998
\end{center}
\begin{abstract}
We analyze string networks in 7-brane configurations in IIB string
theory. We introduce a complex parameter ${\cal M}$ characterizing
equivalence classes of networks on a fixed 7-brane background and
specifying the BPS mass of the network as $M_{BPS} = |{\cal M}|$.  We
show that ${\cal M}$ can be calculated without knowing the particular
representative of the BPS state.  Based on detailed examination of
backgrounds with three and four 7-branes we argue that equivalent
networks may not be simultaneously BPS, an essential requirement of
consistency.
\end{abstract}
\vfill
\setcounter{footnote}{0}
\def\thefootnote{\arabic{footnote}}
\newpage
\renewcommand{\theequation}{\thesection.\arabic{equation}}


\section{Introduction}
\setcounter{equation}{0}

D-branes in superstring theories have recently given new insight into
and powerful tools for studying supersymmetric field theories. The
worldvolume theory of parallel Dp-branes is a p+1 dimensional U(n)
SUSY Yang-Mills, where the Higgs expectation values correspond to the
separation of the branes. The massive gauge bosons arise as open
strings stretched between the D-branes.

7+1 dimensional SYM theories with gauge groups other than U(n) may
also be constructed in IIB string theory
\cite{senorientifold,dasgupta,johansen}, due to the existence of
mutually nonperturbative 7-branes. These branes are labeled by two
coprime integers $[p,q]$ or more precisely by the two-by-two matrix
$({1+pq\;\;p^2 \atop -q^2\;\;1-pq})$, which transforms as a tensor
under the SL(2,Z) duality group. The $[p,q]$-branes serve as endpoints
for $\vect{p}{q}$-strings \cite{wittenbound} and are sources of
monodromies of the axion-dilaton field $\tau$. The single-valuedness
of $\tau$ requires the introduction of branch cuts and as a
consequence, the charges of a general $\vect{r}{s}$-string change upon
crossing the cuts. The configurations with mutually nonlocal 7-branes
give rise to enhanced {\em gauge} groups $SO(n), E_6, E_7$ and
$E_8$. By introducing a D3-brane near the 7-branes, an N=2 SUSY
Yang-Mills theory is realized in its 3+1 dimensional world volume
\cite{bds,senthreebrane}, with the above groups appearing as {\em
global} (flavor) symmetries.

The gauge group of the worldvolume theory
of the 7-branes is not as manifest as it is in pure D-brane
configurations. The states are realized as either
$\vect{p}{q}$-strings between two branes or as multi-pronged string
networks ending on more than two branes \cite{GZ}. The identification
of the gauge group amounts to finding all these BPS string states in a
background with nontrivial metric and SL(2,Z)-monodromy. As an
alternative, one may lift the setup to F-theory \cite{vafa} and deduce
the enhanced symmetry from the singularity of the K3 as the 7-branes
collapse \cite{bikmsv}. 

The string theory picture of the BPS junctions has been subject to
intensive study
\cite{schwarzreview,junctions,sennetwork,rey,krogh,matsuo,callan,kishimoto}
and the emerging picture is quite rich and interesting. It has been
found that the states may correspond to different objects -- strings
or multi-pronged networks -- in string theory, and the realization of
a given state may change as we move around in the moduli space.  The
necessity of three-pronged junctions in the BPS spectrum was first
anticipated by Gaberdiel and Zwiebach \cite{GZ} in 7-brane
backgrounds; the transitions between open strings and junctions were
studied in \cite{GHZ}. Their existence among the ultrashort multiplets
in d=4, N=4 SU(3) SYM as a worldvolume theory of three D3-branes was
showed by Bergman \cite{OB} (see also \cite{threebranes,bergmankol}),
while the role in the spectrum of the d=4, N=2 SU(2) SYM was clarified
in \cite{berfay1,nikita}. String webs also play an important role in
5-dimensional field theories realized as 5-brane webs
\cite{5brane,kolrahmfeld}.

Having anticipated that the multipronged strings are necessary
ingredients of certain IIB configurations the consistency of the
picture demands that the field theory states have unique string theory
representatives. In \cite{GHZ} this was shown in special 7-brane
configurations with constant $\tau$ background and purely conical
metric singularities, by studying the transitions between strings and
junctions. Partial results for general $\tau$ were also reported
showing that a string may not be simultaneously BPS with a junction
which is created from it by a single 7-brane crossing. The aim of the
present notes is to extend the above reasoning to arbitrary 7-brane
configurations and string networks, and argue that a given BPS state
is realized by a unique object at any point in the moduli
space. We believe that our results are useful both in 7-brane theories
and in 3-brane theories in 7-brane background.

We will address the above problem in two steps. First, one should
identify the strings and junctions which -- on the basis of charge
assignments -- are thought to give rise to the same state in the brane
field theory. These are related to each other by a series of crossing
transformations, the process when an open brane (in our case a string)
is created between two branes as they cross each other. (For an
exhaustive classification of the equivalence classes, see
\cite{oliver}.) The determination of the equivalent networks is based
on the monodromies of the axion-dilaton field, but is unaffected by
the actual positions of the branes. We take advantage of this fact by
defining the notion of a {\em graph} which refers to the topology and
homotopy of the object. As the second step, the graphs are mapped to
the concrete spacetime and we may ask the whether this map results in
a geodesic object which possesses the BPS property. One of the
purposes of the paper is to argue that two graphs in the equivalence
class of an open string may not give rise to BPS networks
simultaneously. We will define a complex parameter, ${\cal M}$ which
depends on the equivalence class under consideration but not on the
particular representative and show that the mass of the BPS network is
$|{\cal M}|$. One can also express the slopes of the constituent
strings of a network in terms of ${\cal M}$ which will be an important
ingredient in the argument for the uniqueness of the BPS
representative.

The physics we are considering also possesses a description in terms
of F-theory and M-theory. The strings are lifted to cycles in K3 or
membranes and the BPS condition translates to the requirement that
these membranes be embedded holomorphically, moreover the properties
of the states are determined by the geometry
\cite{wittenphase,leung,kolrahmfeld,bergmankol}.  The uniqueness of
the BPS representative of a state is then seen from the uniqueness of
the holomorphic embedding of the zero-genus curve corresponding to the
open string. (The networks are not unique if they correspond to
higher genus curves). We would like to offer an alternative
viewpoint to this picture; our approach throughout the paper is
conservative in that we restrict ourselves to arguments based on IIB
string theory only.

The plan of the paper is as follows. In \secref{equivalent} we define
the equivalence classes of string networks. In \secref{threesec} we
demonstrate the uniqueness of the BPS object in the simple example of
a background of three 7-branes. In \secref{gensec} we present the
generalizations of the statements of \secref{threesec} which we apply
to the next simplest case of four 7-branes in \secref{foursec}. In
\secref{invariant} we show an explicit construction of the BPS network
generalizing the reconstruction of a geodesic from one of its points
and the slope at that point. In the last section we discuss concrete
transitions in constant $\tau$-background and present an interesting
application showing how the string tensions modify Plateau's problem
on the graphs of least length connecting the vertices of a polygon.

\section{Equivalent graphs}
\label{equivalent}

We consider configurations of parallel 7-branes, whose worldvolume
spans the $x^0\ldots x^7$ directions and as a consequence of this
8-dimensional translational invariance we will restrict ourselves to
the two dimensions of the $x^8x^9$ plane. In the F-theory picture
these two directions are compactified on a sphere and this fixes the
number of 7-branes to be 24, but since we will be interested in
transitions in the close vicinity of groups of branes, we do not need
to care about such ``ultra-global'' behavior of the spacetime. 

Let us first fix the global properties of the background in the
$x^8x^9$ plane. The 7-branes act as sources for the axion-dilaton
field, $\tau$ whose monodromies can described by defining the charges
$[p_i,q_i]$ together with branch cuts emanating from the branes. The
precise position of these cuts is non-physical, but the relative order
is relevant. A typical configuration is shown in \figref{bare}(a);
note that the figure is not meant to indicate any particular position
of the branes in spacetime, only the global properties of $\tau$ are
defined (see \cite{GHZ} for the precise definition of the covering
space). Our conventions are the same as in \cite{GHZ}; the examples
will involve {\bf A}- and {\bf C}-branes whose monodromies are:
\bea
{\bf A}=& [1,0]: \,\,K_A = & T^{-1} = 
\pmatrix{1 & -1 \cr 0 & 1} \,, \nonumber\\ 
{\bf C}=& [1,1]: \,\, K_C = &  T^{2}S 
             = \pmatrix{2 & -1 \cr 1 & 0},\nonumber
\eea  
here a $\vect{p}{q}$ string becomes a $K\vect{p}{q}$-string upon
crossing the corresponding cut in the anticlockwise direction. 
\onefigure{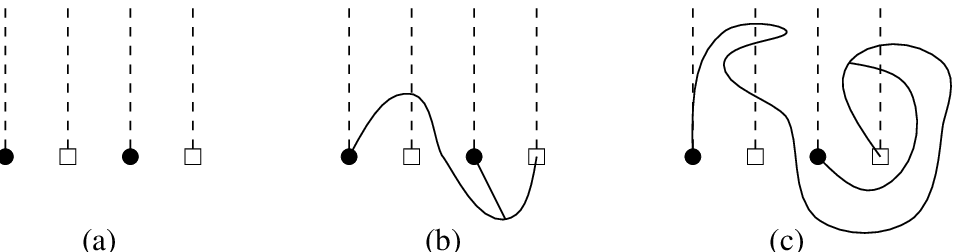}{(a) We first define the monodromies of $\tau$. The
filled circles stand for {\bf A}-branes, the empty squares are
{\bf C}-branes. (b) A graph that is allowed by charge conservation. (c)
The same graph as in (b).} 

As far as the homotopy of curves is concerned any string or network
can be represented within the above figure. We define a {\em graph} to
be a collection of curves joining three-pronged junctions and/or
branes. An example is presented in \figref{bare}(b) where three curves
join a three-pronged junction with three 7-branes.  We do not allow
for four- or more-pronged junctions, these will arise as limiting
cases of the objects described by our graphs.  The curves in a graph
represent $\vect{p}{q}$-strings and thus charge conservation restricts
the allowed graphs. In \figref{bare}(b) and in most cases that we
shall consider, the charges are uniquely determined (up to an overall
sign) by the homotopy and we will not indicate them. We stress that
the notion of a graph does not refer to the actual paths or angles in
spacetime thus there is no distinction between the graphs in
\figref{bare}(b) and (c): they are said to be equal. Also we consider
two graphs equal if one is obtained from the other by pulling a
$\vect{p}{q}$-curve through a $[p,q]$-brane as a string does not feel
monodromy around a brane it may end on.

Two graphs will be said to be {\em crossing-transformed} of each other
if one is obtained from the other by pulling one of the strings
through a brane and creating an extra prong \cite{hananywitten} as for
example in \figref{hw}. See \cite{GZ,GHZ} for a detailed description
of the charges and charge conservation. In general the resulting prong
may not have coprime charges.  
\onefigure{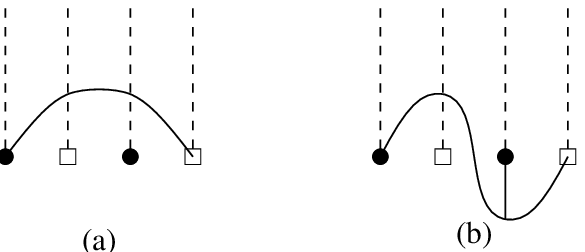}{Crossing-transformation.}

We will call two graphs {\em equivalent} if they are related by a
series of crossing-transformations and channel-transitions, the latter
corresponds to changing a ``t-channel subdiagram'' to an ``s-channel
subdiagram'' without altering the outgoing prongs, see
\figref{crossing}.  
\onefigure{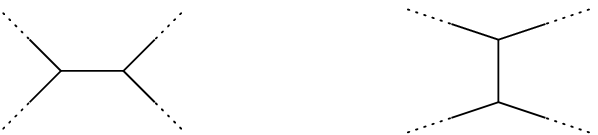}{Channel transition.}

A {\em generalized string} or {\em network} is a continuous map of a
graph onto $S^2$ with the obvious condition that the ``branes'' are
mapped to the corresponding branes and the homotopy is preserved. Two
junctions are said to be equivalent if their graphs are equivalent. In
this paper we shall restrict ourselves to networks which are
equivalent to an open string; these do not have moduli parameters and
will be uniquely realized. 

We will be looking for graphs that are BPS in the background of the
7-branes. We will mostly rely on two necessary conditions
\cite{sennetwork}: in a BPS string network the strings are smooth
geodesics and the tensions balance each other at the junction
points\cite{schwarzreview}.

\section{BPS state in the vicinity of three 7-branes}
\label{threesec}

As a warm-up we restrict ourselves to the vicinity of three
branes. This situation in the background of constant axion-dilaton
field was discussed in \cite{GHZ}. We will be interested in networks
whose graphs are equivalent to that of a given open string, the one
shown in \figref{three}(a). Application of crossing transformations
through the various branes results in the equivalent graphs shown in
\figref{three}(b),(c), and (d). We shall see that the endpoints of a
graph determine the BPS network (independent of the multiplicity of
the prongs), thus as far as the possible endings
are concerned, the four graphs form an exhaustive list.  Notice that
it is unnecessary to extend the list with graphs which have both
incoming and outgoing prongs on the same brane. The reason is that
these pairs of prongs are not fixed to the brane, they can be
considered as parts of a string passing by; \figref{doubleprong} shows
a network equivalent to the ones in \figref{three}.
\onefigure{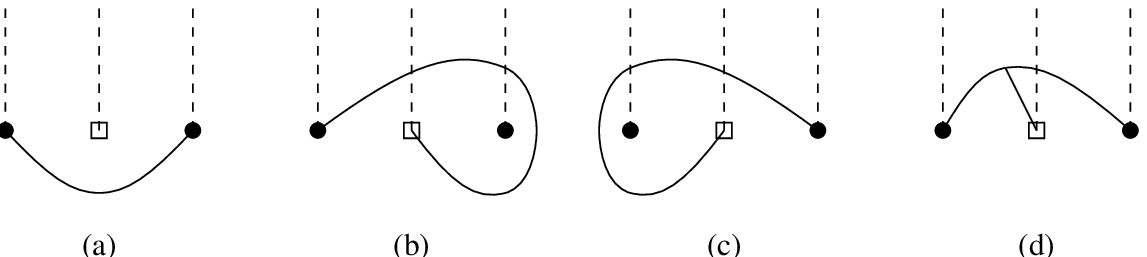}{Four equivalent graphs in the vicinity of three
branes.}
\onefigure{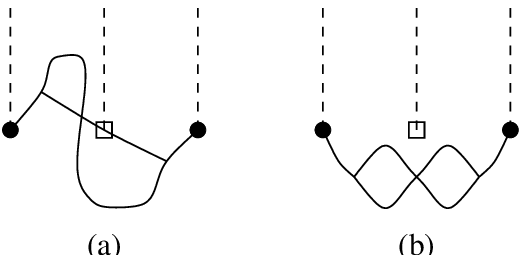}{(a) A network with a pair of incoming and
outgoing strings on the middle brane. (b) A deformation of the previous
network shows that the middle brane does not play an important role
and we obtain an {\bf A-A} string with a closed loop in the
middle. This can never be BPS as will be shown in \secref{foursec}.}

Depending on the positions of the three 7-branes a given graph of
\figref{three} may or may not be realized as a BPS object. In the
following we will show that there is no point in the moduli space
where any two of the above graphs may be BPS.  

Let us  assume that for any of the above four graphs there is a region
in the moduli space where a BPS state homotopic to the graph
exists. Based on this assumption we shall determine the BPS mass and
the slopes of the geodesics at the branes. We will find that this data
will be the same for the four graphs. 

We choose $z$ as the complex coordinate on the cut plane and the
positions of the branes are denoted by $P_1$, $P_2$ and $P_3$ (see
\figref{p1p3}). Let the 7-branes at $P_1$ and $P_3$ be $[p,q]$-branes
and the one at $P_2$ an $[r,s]$-brane and also define $e$ as
\bea
e&=& ps-qr.
\eea
Note that the charges of the 7-branes are now arbitrary, they need not
be {\bf A}- and {\bf C}-branes.
\onefigure{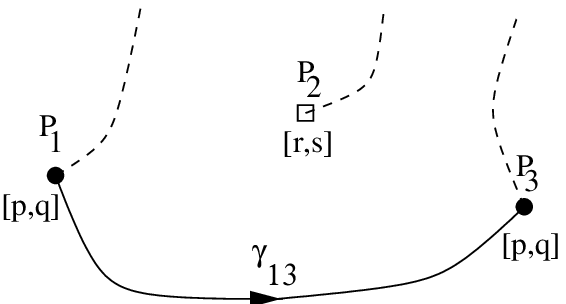}{Direct string corresponding to
Fig.~\protect\ref{three}. When there is a BPS state represented by
this graph its mass is given as an integral along $\gamma_{13}$. }

The mass of a $\vect{p}{q}$ string can be calculated using the
tension-weighed metric:
\bea
ds^2_{\vect{p}{q}} &=& T_{\vect{p}{q}}d^2s =
|h_{\vect{p}{q}}(\tau(z),z)dz|^2 
\eea
where $h_{\vect{p}{q}}(z)$ and $\tau$ is given in terms of the
position of the branes, $z_i$ as \cite{greeneetal}:
\bea
\label{metric}
h_{\vect{p}{q}}(\tau(z),z)&=& (p-q\tau(z))
\eta^2(\tau)\prod_i(z-z_i)^{-\frac{1}{12}} \\
j(\tau(z)) &=& \frac{4(24f)^3}{\Delta}; \;\;\;\;\;\;\;
\Delta \equiv 27g^2+4f^3 = \prod_{i=1}^{24}(z-z_i).
\eea
When the direct string -- which is represented in \figref{three}(a) and
\figref{p1p3} -- runs along a geodesic, its mass is given by:
\bea
m &=& |{\cal M}_{(a)}| \\
{\cal M}_{(a)} &=& \int_{\gamma_{13}}h_{\vect{p}{q}}(\tau(z),z)dz,
\label{Mdef}
\eea
where the path of the integration, $\gamma_{13}$ goes from $P_1$ to
$P_3$ and below $P_2$ as is shown in the figure, but is otherwise
arbitrary, in particular it need not run along 
the geodesic. To simplify most of the formulas we introduced the
complex quantity, ${\cal M}_{(a)}$ whose absolute value is the mass of
the BPS state. 

Now we determine the slope of this geodesic at its left endpoint,
$P_1$. Since the metric is locally flat, in the coordinate  
\bea
w &\equiv& \int_{P_1}^z h(\tau(z),z) dz,
\label{wdef}
\eea
the geodesic is a straight line. In the above integral $h(z)$
is the instantaneous metric that the string feels along its path which
in this case equals  $h_{\vect{p}{q}}(z)$. 
The slope of the string at a generic point, $z$ is given by:
\bea
\arg(dz) &=&
\arg\left(\frac{dw}{\frac{dw}{dz}}\right)  = 
\arg\left(\frac{{\cal M}_{(a)}}{h(z)}\right),   
\label{slopedef}
\eea
and thus the initial slope at $P_1$ is equal to:
\bea
\mbox{slope}_{P_1} = \arg\left(\frac{{\cal M}_{(a)}}{h(P_1)}\right).
\eea
Note that in this formula $h(P_1)$ is $h_{\vect{p}{q}}(P_1)$,
which is nonsingular at a $[p,q]$-brane as follows from \myref{metric}.

When the arrangement of the branes gives rise to a BPS state
represented by the graph of \figref{three}(b) the mass can be
computed similarly but one must note that for the above arrangement of
branch cuts this is a string with nonuniform
charges. Nevertheless the mass is still given as length times tension:
\bea
{\cal M}_{(b)} &=& \int_{P_1}^{C_2}h_{\vect{p}{q}}(\tau(z),z)dz +
\int_{C_2}^{C_3}h_{\vect{p-er}{q-es}}(\tau(z),z)dz + \nn \\
&&+\int_{C_3}^{P_2}h_{\vect{-er}{-es}}(\tau(z),z)dz.
\label{indmass}
\eea
The integration is performed along the curve $\gamma_{12}$ which is
depicted in \figref{p1p2}(a) and represents a $\vect{p}{q}$-string
between $P_1$ and the $C_2$ cut followed by a
$\vect{p-er}{q-es}$-string between $C_2$ and $C_3$ and a
$\vect{-er}{-es}$ which ends on $P_2$ 
\onefigure{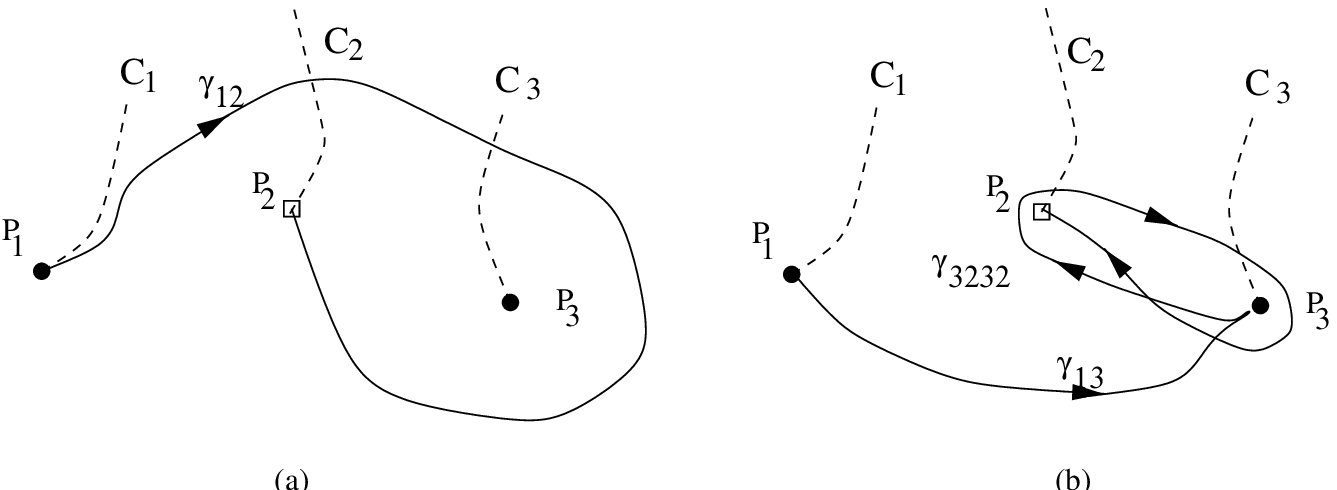}{(a) The curve $\gamma_{12}$ which is used to compute
the mass of the BPS indirect string. (b) We transform $\gamma_{12}$ 
into $\gamma_{3232}\circ\gamma_{13}$ to benefit from that the integral
over $\gamma_{3232}$ vanishes. }

Let us now deform the curve into a special one so that the integral in
\myref{indmass} becomes simple. The new path is shown in
\figref{p1p2}(b): it first goes along $\gamma_{13}$ which hits $P_3$,
then loops around $P_2$ and then $P_3$ following $\gamma_{3232}$
which ends on $P_2$. The integral along $\gamma_{3232}$ can be
evaluated by shrinking the path into a single curve between $P_3$ and
$P_2$ which is covered by $\gamma_{3232}$ three times with different
charges:
\bea
\int_{\gamma_{3232}}h(\tau(z),z)dz &=&
\int_{P_3}^{P_2}  \left(
h_{\vect{p}{q}}(\tau(z),z)-
h_{\vect{p-er}{q-es}}(\tau(z),z)+ \right. \nn \\
&&\left. \;\;\;+h_{\vect{-er}{-es}}(\tau(z),z) \right) dz = 0,
\eea
by the linearity of the metric in the charges. 
The vanishing of this integral implies that ${\cal M}_{(b)}$ is given
by the integral over $\gamma_{13}$ only and is equal to
${\cal M}_{(a)}$. We again define $w$ as in \myref{wdef} keeping in
mind that the instantaneous metric is not $h_{\vect{p}{q}}$ along the
whole path, nevertheless it is analytic because of the
SL(2,Z)-invariance at the branch cuts. The slope at $P_1$ is again
given by 
\bea
\mbox{slope}_{P_1} = \arg\left(\frac{{\cal M}_{(b)}}{h(P_1)}\right) =
\arg\left(\frac{{\cal M}_{(a)}}{h(P_1)}\right).
\eea

Entirely similar argument holds for the indirect string of 
\figref{three}(c) and we conclude that:
\bea
{\cal M}_{(a)} = {\cal M}_{(b)} = {\cal M}_{(c)} =
\int_{\gamma_{13}}h_{\vect{p}{q}}(\tau(z),z)dz.
\eea

Now let us turn to the three-pronged junction of \figref{three}(d),
which was discussed previously \cite{GHZ}. The calculation of ${\cal
M}_{(d)}$ is simplified if we use the freedom of relocating the $C_2$
cut which we place right on top of the $P_2$-prong as is shown in
Fig~\ref{p1p2p3}(a) \onefigure{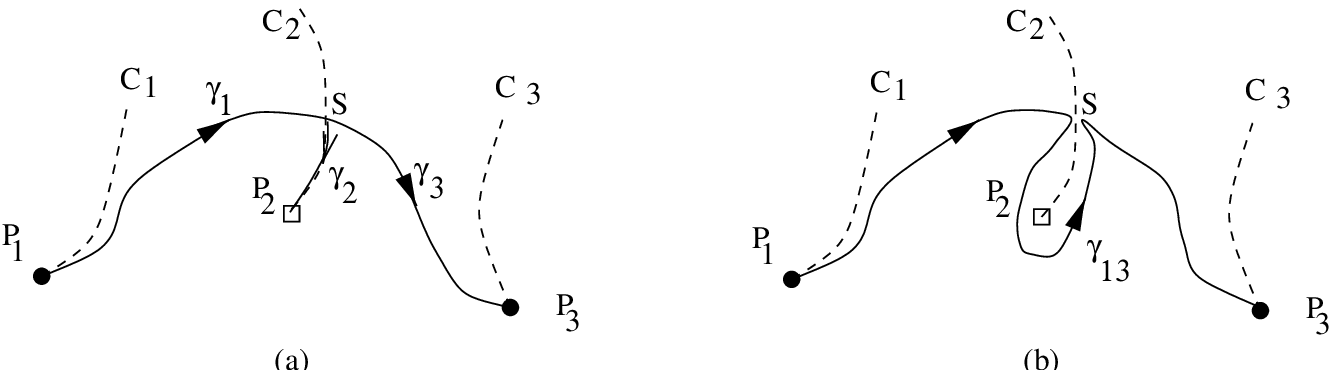}{(a) The three-pronged junction
of Fig.~\protect\ref{three}(d) with a convenient placement of the
$C_2$-cut. (b) The length of the $P_2$-prong is equal to the integral
over the loop around $P_2$.}

The junction point which is also on the $C_2$-cut, is denoted with
$S$. Our claim is that the tension-weighed integral along the
$P_2$-prong is actually equal to the integral along the loop which
departs and arrives at $S$ and surrounds $P_2$ weighed with the
tension of the $\vect{p}{q}$-string (\figref{three}).  Let us pause
for a short comment here. This Ansatz is not as ad hoc as it seems at
first sight but is motivated by the physical picture which one has in
mind when visualizing what really happens in the Hanany-Witten effect
considered as a real process. \figref{hwprocess} shows a (non-BPS)
string in the vicinity of a $[1,1]$-brane. (To make the argument more
clear it is now useful to place the cut on the string side -- as
opposed to the junction side in Fig.~7.) The $\vect{0}{-1}$ and
$\vect{1}{0}$ strings attract each other and are pulled towards each
other until they form the $\vect{1}{1}$ bound state. If the other ends
of the strings are fixed, then the bound state can only be formed on a
short segment originating from the [1,1]-brane and we see a junction
as a result.
\onefigure{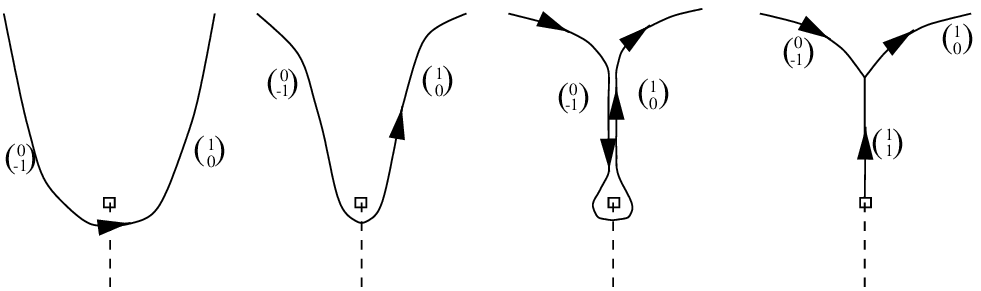}{A junction is formed as a result of the
Hanany-Witten effect.}

Back to Fig~\ref{p1p2p3}(b), the integral is most easily performed
after shrinking the loop to a curve between $S$ and $P_2$ which is
covered twice:
\bea
\oint_S^S h_{\vect{p}{q}}(\tau(z),z)dz &=& 
\int_S^{P_2} \left(h_{\vect{p}{q}}^{left}(\tau(z),z) - 
h_{\vect{p}{q}}^{right}(\tau(z),z)\right) dz = \nn \\
&=& -e\int_S^{P_2} h_{\vect{r}{s}}(\tau(z),z)dz, 
\label{pronglength}
\eea
where we used that the limiting values of the metric on the two sides
of the cut relate to each other as $h_{\vect{p}{q}}^{right} =
h_{M^{-1}_{r,s}\vect{p}{q}}^{left} = h_{\vect{p+er}{q+es}}^{left}$. 
If we now define the quantity analogous to ${\cal M}_{(a)}$,  
\bea
{\cal M}_{(d)} &\equiv& \int_{\gamma_1} h_{\vect{p}{q}}(\tau(z),z) dz +
(-e \int_{\gamma_2} h_{\vect{r}{s}}(\tau(z),z) dz) + \nn \\
&& + \int_{\gamma_3} h_{\vect{p}{q}}(\tau(z),z) dz, 
\label{massd}
\eea
then we again learn that ${\cal M}_{(d)}={\cal M}_{(a)}$.  Let us
now show that the mass of the geodesic object is again $m = |{\cal
M}_{(d)}|$. The absolute value of each of the three complex terms 
in \myref{massd} is equal to the mass of the corresponding geodesic
prong, therefore the sum of them gives the mass of the junction. For
a general placement of the junction point $S$, this mass will be
greater than the BPS bound. This bound -- which equals the mass of the
BPS three-pronged string -- is easy to find: the sum of the absolute
values of the three complex numbers is greater then the absolute value
of the sum,
\bea
m \ge |{\cal M}_{(d)}|,
\eea
and is minimal, when they
are parallel on the complex plane, and exactly in this case the total
mass becomes:
\bea
m_{\mbox{\tiny geod.junction}} = |{\cal M}_{(d)}|.
\eea
In \cite{GHZ} we pointed out a geometrical interpretation of this
result: in the coordinate
$w_{\vect{p}{q}}(z)\equiv\int^zh_{\vect{p}{q}}$, the junction can be
represented as a broken line segment between $P_1$ and $P_2$.  This
crosses the image of the $P_3$-cut, which shows up as an excised
region on the $w$-plane. Eq. \myref{pronglength} shows that the length
of the $P_3$-prong equals the line segment inside the excised region
and it is clear that the total length is minimal when the three parts
of this broken segment are parallel, which is equivalent to the above
result. Also, in the vicinity of the junction point $S$ the fact that
the three complex numbers in \myref{massd} have equal argument
translates to the balance of forces \cite{sennetwork}.

Now we calculate the slope at $P_1$, similarly to the previous
cases. We find that 
\bea
\mbox{slope}_{P_1} &=& \arg\left(\frac{w(S)-w(P_1)}{h(P_1)}\right) =
\arg\left(\frac{\int_{\gamma_1} h_{\vect{p}{q}} dz}{h(P_1)}\right) =
\nn \\
&=& \arg\left(\frac{{\cal M}_{(d)}}{h(P_1)}\right) = 
\arg\left(\frac{{\cal M}_{(a)}}{h(P_1)}\right).
\eea
The calculation can be repeated for the other two endpoints.

We have learned that the four quantities ${\cal M}_{(a)}$, ${\cal
M}_{(b)}$, ${\cal M}_{(c)}$ and ${\cal M}_{(d)}$ whose role is to
define the mass of the BPS state in a certain regime of the moduli
space, are equal, independent of the position of the branes. This
shows that the mass of a BPS state is defined unambiguously, no matter
which of the four integrals is used and is independent of the actual
graph which represents the BPS state. Moreover, whichever graph is
realized as a geodesic object, its slopes at the endpoints are given
in terms of the quantity ${\cal M}_{(a)}$ and $h(P_i)$ which depend
only on the brane configuration.

Now it follows that there is no 7-brane configuration where two of
these equivalent graphs would simultaneously be realized as BPS
states. To see this recall that a geodesic curve is fully determined
by one of its points and the slope at that point. Take for example the
two strings of \figref{three}(a) and (b), these have a common endpoint
at $P_1$. If the two strings were both BPS, their slope at $P_1$ would
be equal which would imply that the two geodesics are identical. In
other words, the slope at $P_1$ is given in terms of the metric and
the string ``decides'' which other brane it wants to end on. Let us
now turn to \figref{three}(d) and compare it to one of the open
strings, say (a). They have two common endpoints on the {\bf
A}-branes, where the initial slopes are determined. One can construct
both of the entire geodesics based on the initial data. It is now
clear that they either coincide and give rise to the open string or
intersect each other at an angle, which leads to the three-pronged
network but can not do both; thus there may not be any configuration
when both a string and a junction is BPS. This is again clear from the
geometrical interpretation\cite{GHZ}: the line segment between two
branes on the $w$-plane either avoids or crosses the excised region of
the cut of the third brane; in the previous case the BPS object is a
string while in the latter case it is a junction.

\section{General properties of equivalent graphs}
\label{gensec}

In this section we would like to outline those lessons from the case
of three 7-branes which generalize to more complicated graphs. We will
not present a rigorous general proof for the existence and uniqueness
of the BPS representative of equivalence classes, but we believe and
our experience shows that these properties are sufficient to analyze
concrete configurations.

The main result of the previous section was that we found certain
relevant quantities which characterize an {\em equivalence class} of
the networks and depend only on the details of the background but not
on the particular representative. We assigned a complex number, ${\cal
M}$ to an open string and observed that it was not changed as we
performed crossing transformations. We emphasize that this does not
mean that ${\cal M}$ is invariant as we move the the 7-branes and the
BPS object changes its shape, but rather that at a given point in the
moduli space one can determine this number without knowing the
representative of the equivalent networks to which it belongs. The
central idea -- that ${\cal M}$ does not change when a crossing
transformation is made -- is based on the observation that the
integral over the prong which was created (removed) equals an integral
over the string that surrounds the brane, see
eq.\myref{pronglength}. Note also that the statement is not restricted
to the cases when a single prong is created as $e=ps-qr$ could be any
integer. This holds independently of the particular background and the
network that this prong is part of; thus we conclude that ${\cal M}$
is a parameter of equivalence classes of string networks and the mass
of the BPS representative of the given class is:
\bea
m_{\mbox{{\tiny BPS}}}=  |{\cal M}|
\eea

We also discussed the slopes of the strings in a network and found
that at any endpoint the slope of the string is independent of the
representative. What we showed was that the slope can be calculated in
terms of ${\cal M}$ and thus is invariant under crossing
transformation. According to the previous paragraph, this is again a
general result, moreover nothing relied on that the point where we
calculate the slope is the endpoint of a string, thus we arrive at the
following conclusion. The slope of a $\vect{p}{q}$-string at point $z$
on the $x^8x^9$-plane, in a BPS network which belongs to the
equivalence class characterized by ${\cal M}$ is:
\bea
\mbox{slope}_{\vect{p}{q}}(z) &=& 
\arg\left(
\frac{{\cal M}}{h_{\vect{p}{q}}(z)}\right)
\eea
In the following section we will show how these results can be applied
to concrete situations.

\section{An example with four 7-branes}
\label{foursec}

In section \ref{threesec} we asked what happens when a mutually
nonperturbative 7-brane appears in the vicinity of two D-branes, where
these two alone would correspond to a spontaneously broken
SU(2)-theory. In the 7+1 dimensional worldvolume of the three 7-branes
it is still an SU(2) gauge theory, however the W-boson is no longer
simply an open string between the two [1,0] branes, rather --
depending on where we are in the moduli space -- it is represented by
one of four equivalent graphs.

If we add more branes, we may get other gauge groups and the
interesting SO(8) and exceptional symmetries can be realized.  When we
consider these theories with more and more 7-branes, both the number
of equivalence classes and the number of equivalent graphs in a given
class increase dramatically. \cite{oliver} gives a systematic
approach to finding the equivalent networks and also provides the
classification of equivalence classes in terms of representations of
the underlying gauge symmetry. Enumerating the equivalent graphs for
many 7-branes is complicated and we will see that the vicinity of
four branes already gives a great deal of insight into the problem.

The particular setup that we shall now investigate is that of
\figref{bare}(a) which corresponds to a resolution of a singularity
possessing the SU(3) gauge group. This corresponds to a fiber of type
IV in the elliptic fibration. The simplest representatives of the
W-bosons corresponding to the positive roots are shown in
\figref{su3wboson}(a), (b) and (c). 
\onefigure{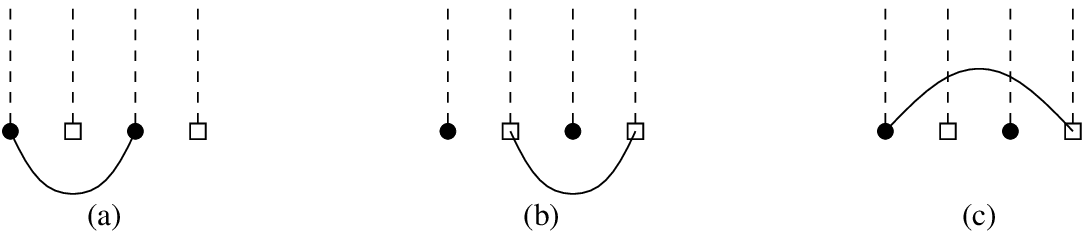}{The three inequivalent strings that are in
one-to-one correspondence with the positive roots of SU(3).}
The filled circles stand for [1,0]-branes and the empty squares for
[1,1]-branes. We shall study the state of \figref{su3wboson}(a), 
for which a family of equivalent junctions is shown in \figref{four}.
\onefigure{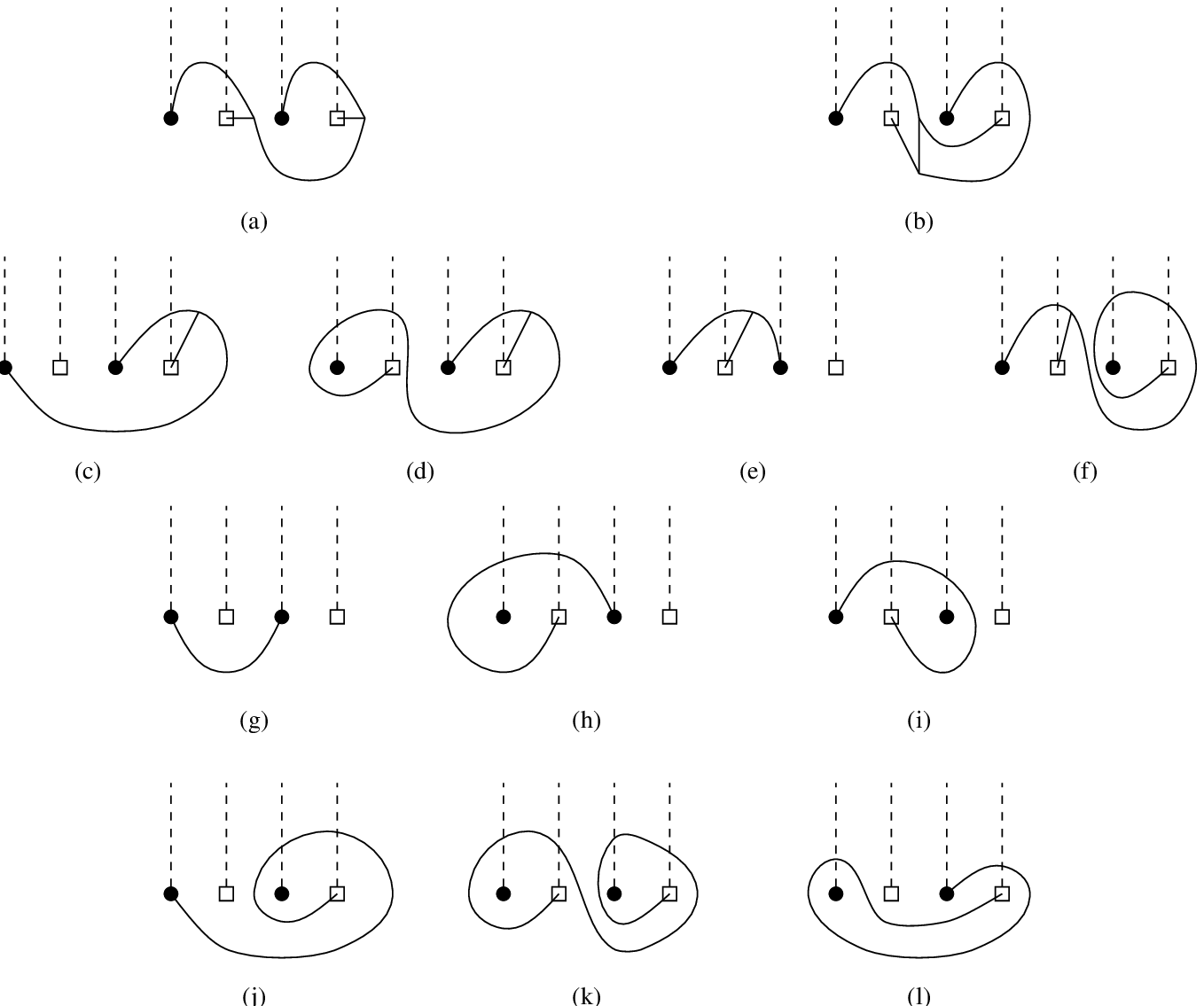}{Twelve equivalent graphs in the vicinity of four
7-branes.}

As we can see, six equivalent strings are listed together with four
different three-pronged junctions and two four-pronged ones, the
latter differing from each other by a channel-transition. The
charges for any of the open strings can be read off from the branes
and are determined for the other objects by noting that they are
related by crossing-transformations.

The graphs of \figref{four} -- when realized as BPS networks --
correspond to a particular $W$-boson of the SU(3)-model. The
consistency of the low energy theory requires that there be a unique
representative of this state. We shall now demonstrate -- using the
arguments of section \ref{gensec} -- that this is the case. Since we
do not have a general proof in hand, we do this by pairwise comparison
of the members of this class; that is we show that no two graphs may be
realized simultaneously as the BPS state. First, recall that at a
given point in the moduli space the complex number ${\cal M}$ is a
parameter of the whole class and also that whichever network is
realized, the slope of a string at a given point $z$ is determined by
its charges and ${\cal M}$.

First of all, none of the four-pronged networks, (a) and (b) may
``coexist'' with any of the other graphs: if say (a) and (e) were both
BPS, that would be in conflict with the fact that the slopes of the
three prongs ending on the three leftmost brane (which are given in
terms of ${\cal M}$) determine the whole graph of (e) and there is no
fourth prong. (This is analogous to the argument in \secref{threesec}
about the graphs of \figref{three}(a) and (d)). Notice also that there
are groups of graphs where one of the four branes has no prong on it,
like (c)-(g)-(j)-(l), (d)-(h)-(k)-(l), (e)-(g)-(h)-(j) and
(f)-(i)-(j)-(k).  For these the arguments of \secref{threesec}
regarding the absence of simultaneous BPS junctions can be repeated
without modification, because they did not depend on how the nearby
7-branes affect the metric. (One should not worry about their cut
either as it can be transformed away by an SL(2,Z)-transformation
\cite{GHZ}.)  Now consider any two of the six open strings of (g)--(l)
and notice that they necessarily have a common point where both have
the same charges. At this point on the $z$-plane, the slope of the
string is given in terms of ${\cal M}$ which can not give rise to two
different geodesics. What about the pairs like (e) and (j), a
three-pronged network and a string with one common endpoint? The
string is certainly determined in terms of its slope on the common
brane but it is not clear that the three-pronged junction cannot be
simultaneously BPS with one of its prongs on top of the (j)-string as
in \figref{coexist}.
\onefigure{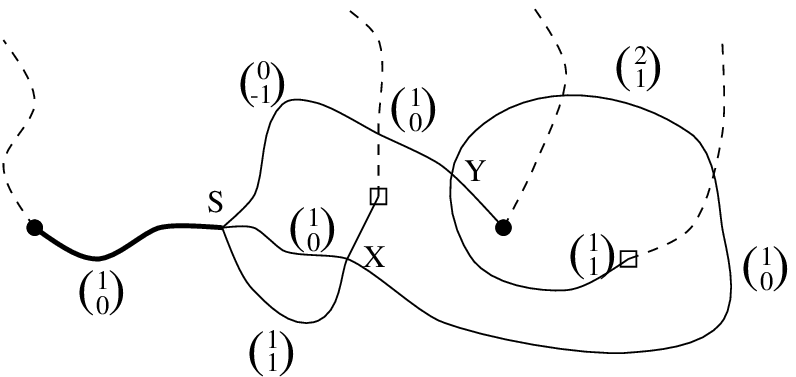}{Comparing the equivalent string of
                    \figref{four} (j) and three-pronged network (e) on
                    the $z$-plane.}

Note that at the intersection point $Y$, the slope argument does not
work, since the charges of the two objects are different. However, the
two networks can not both be BPS between $S$ and $X$ as follows from
the following simple argument.  Consider the $\vect{1}{0}$ and any
$\vect{p}{q}$ strings in a similar configuration depicted in
\figref{sen}.  
\onefigure{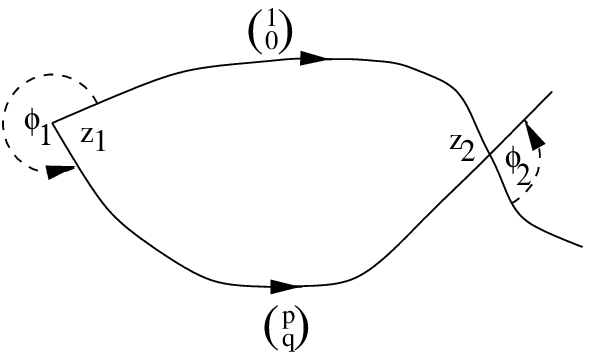}{Two strings whose charges are different intersecting
twice. This configuration can never be BPS as follows from the
condition of the angles.} 
If the $\vect{1}{0}$ and $\vect{p}{q}$ geodesics intersect at two
points, one of the angles, $\phi_{i}$ at the intersection points
have to fall between $0$ and $\pi$ while the other should be between
$\pi$ and $2\pi$. Supersymmetry is maintained at the two intersection
points if the angles satisfy \cite{sennetwork}:
\bea
\phi_{i} &=& \arg(p-q\tau(z_{i})), \;\;\; \mbox{or} \nn \\
\phi_{i} &=& \arg(p-q\overline{\tau}(z_{i})),
\eea
which however would imply that $\Im\tau$ is positive at one of the
intersection points and negative at the other, but the latter is not
possible.  Having ruled out this last possibility we conclude that the
BPS representative of the equivalence class in \figref{four} is
unique.

The above analysis can be repeated without difficulty for the other
two equivalence classes of \figref{su3wboson}(b) and (c). 

\section{Invariant charges and construction of the BPS network}
\label{invariant}

To find the entire trajectory of a BPS string it is enough to
determine a point of a curve with the slope at that point. Starting
from that point one can follow the geodesic path until the 7-brane
where the string ends is hit. The argument in connection with
\figref{coexist} suggests a more general feature: it seems that the
general network can be constructed similarly from the same data, even
when it is more complicated than a string. We did not prove this
conjecture in general but will now present the idea of this explicit
construction, which has been applicable to every concrete case we
studied.

To understand the construction we need to know the concept of
invariant charges assigned to the 7-branes. This is explained in
detail in \cite{oliver}, we summarize here the main points only. Given
a 7-brane background and a graph, we can assign to every brane an
integer which is expressed in terms of the number of prongs ending on
that brane and the number of prongs crossing its branch cut. The
definition of this invariant charge is such that it does not change in
crossing transformations and thus corresponds to the equivalence class
of the networks and not the particular graph. 

Let us return to the simplest case of the configuration with three
branes and take the equivalence class of \figref{three}. The invariant
charges of the leftmost, middle and rightmost branes are +1, 0 and -1,
respectively when the orientation of $\vect{1}{0}$-string points from
the left to the right. Let us assume that our BPS object starts on the
leftmost brane and try to construct it. The slope at the initial point
is given so we simply have to follow the $\vect{1}{0}$-geodesic on the
$z$-plane. Now there are two possibilities: this geodesic either
crosses the cut of the middle brane or it does not. In the latter case
it simply hits the rightmost brane and we are done.  (It cannot miss
it as follows from \myref{Mdef} and \myref{slopedef}.)  
\onefigure{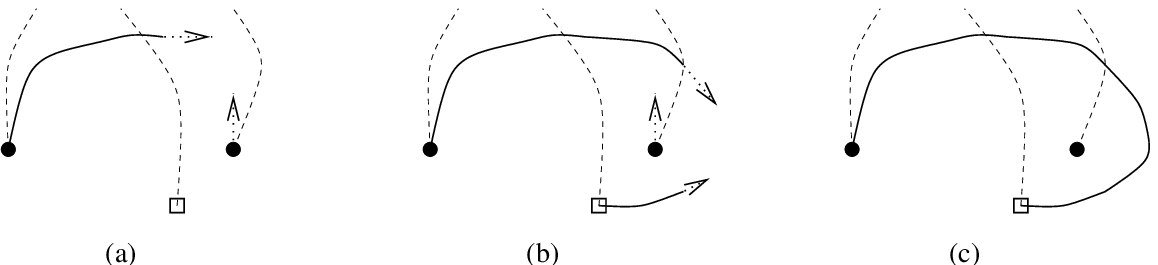}{Construction of the BPS indirect string. (a) The
geodesic string hits the branch cut of the [1,1]-brane. (b) To restore the
invariant charge of the [1,1]-brane a prong is started with the given
initial slope. (c) Because the string crossed the branch cut of the
rightmost brane, no prong should end on that brane to preserve its
invariant charge. The geodesics of the leftmost and middle brane meet
smoothly.}
\onefigure{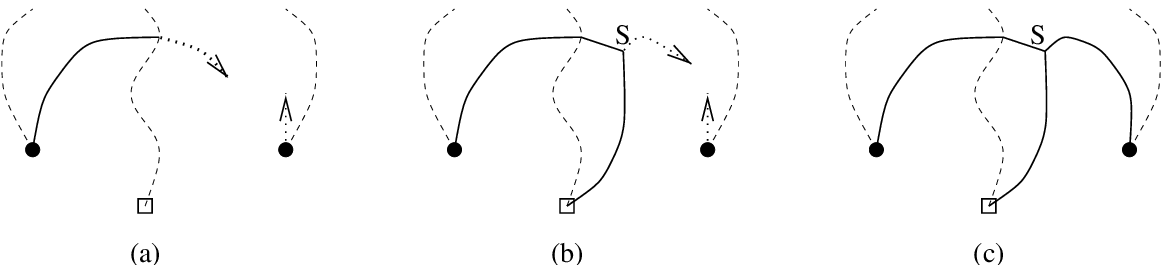}{Construction of the BPS junction. (a) The
geodesic string hits the branch cut of the [1,1]-brane. (b) To restore the
invariant charge of the [1,1]-brane a prong is started with the given
initial slope which meets the previous geodesic at the junction
point. (c) The initial slope of the third prong determines the third
geodesic ending on the rightmost brane.}

However, if the geodesic crosses that cut then the invariant charge of
the middle brane is temporarily changed which we should compensate by
starting a prong (with the slope given in terms of ${\cal M}$) from
the brane. This new prong again runs along its geodesic which we can
follow until it meets our first prong. Now there are again two
possibilities: they may meet smoothly such that they together form a
geodesic and we end up with \figref{four}(b). Note that in this case
the charges work out only if the cut of the rightmost brane is
crossed, when we meet that cut with either geodesic, we change the
invariant charge of the rightmost brane which is compensated by
erasing its prong, see \figref{construct2}.  If the two geodesics meet
under an angle at a point $S$, then a third geodesic should be started
with the proper charges and slope from the junction point which
eventually hits the third brane, see \figref{construct1}.  Note that
it is guaranteed by supersymmetry that the two geodesics meet under
the proper angle to form the BPS junction.

We used the simplest configuration to illustrate our idea of how an
initial point, the slope of the string at that point and the set of
invariant charges determine the whole BPS object. We believe that the
recipe that we gave above similarly determines BPS networks in more
complicated backgrounds.

\section{Four-pronged network in constant $\tau$}
\label{constant}

Looking at some of the networks in \figref{four} one could say that we
might overestimate the complicatedness of the background and though
charge conservation permits all, some of the graphs will never be
BPS. This question is hard to answer because the explicit form of the
metric is indeed very complex and it is unlikely that an analytic
determination of the geodesics is possible, though a numerical
analysis would be desirable and useful. We will not do that, but will
argue that the moduli space of graphs is not less complex than
suggested by \figref{four} using a similar but far more tractable
configuration.

The idea is borrowed from \cite{GHZ} where we used configurations of
7-branes with constant axion-dilaton field. This can be achieved by
collapsing some of the branes; in particular $\tau =
e^{i\frac{\pi}{3}}$ can be maintained if the background consists of
coinciding pairs of [1,0]- and [1,1]- branes. Thus we may get some
insight into the moduli space of the equivalence class of
\figref{four} if we put a [1,0]-brane on top of each [1,1] and a
[1,1]-brane on top of each [1,0] as shown in \figref{e6const}. This is
of course no longer the general resolution of an SU(3) singularity,
but rather a special resolution of an $E_6$ configuration (see
sec. 3. of \cite{GHZ}).
\onefigure{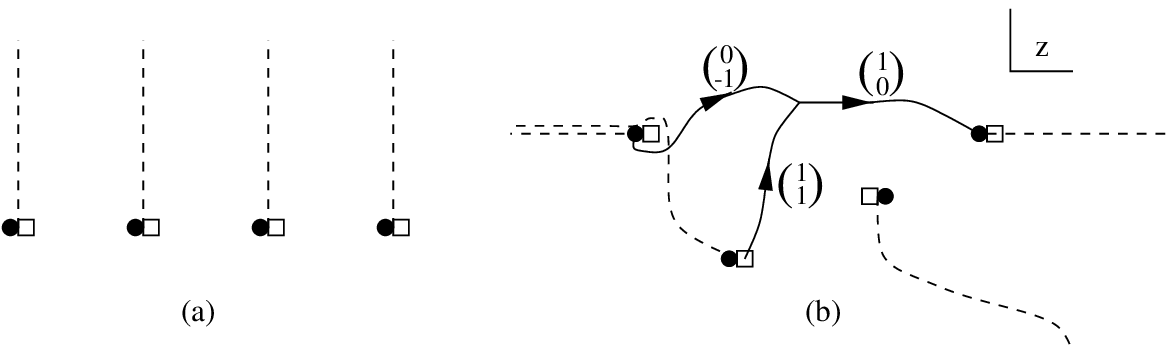}{(a) Constant $\tau = e^{i\frac{\pi}{3}}$
resolution of th $E_6$ singularity. (b) A convenient arrangement of
the cuts used later in \figref{metamorph}, and a three-pronged
network.} 
\onefigure{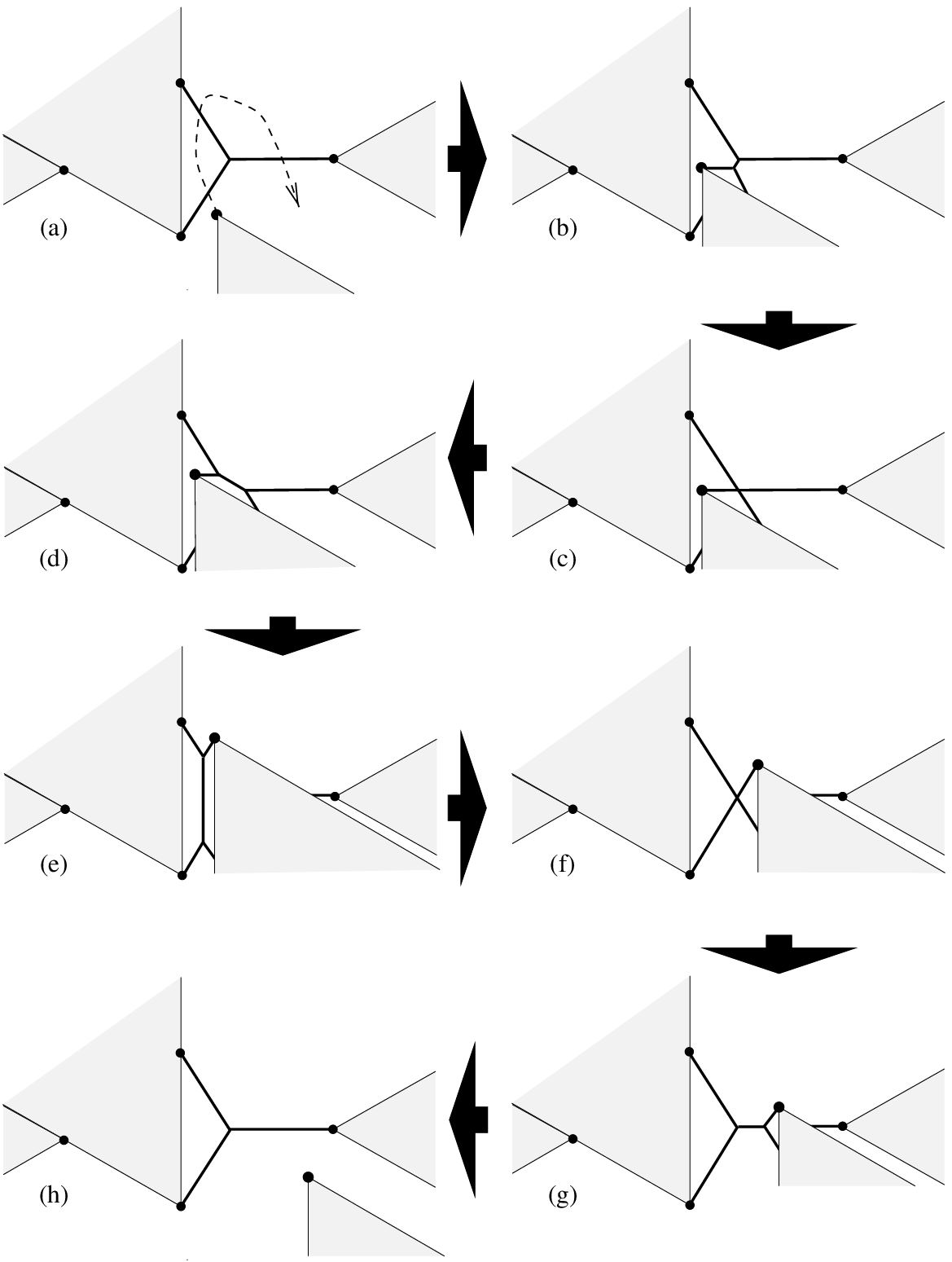}{Metamorphosis of a four-pronged network}

This configuration possesses a much simpler metric than a general one:
in the $w$-coordinate (see \myref{wdef}) {\em all strings} are
straight lines and the four singularities are purely conical, giving
rise to an excised region of deficit angle $60^\circ$ emanating from
each puncture. Now it is just an exercise in planar geometry to show
that every single graph of \figref{four} is realized at certain
regions of the moduli space. We leave it to the reader to confirm that
the moduli space is an extended version of Fig.~13 of \cite{GHZ} and
we only present an interesting part of it in \figref{metamorph}
corresponding to the metamorphosis of the four-pronged junctions.

It is reassuring to see that not only the crossing transformations but
also the channel-transitions are smooth processes. It is instructive to
see this in more detail.  We show in \figref{plato} the
(b)$\longrightarrow$(d) process of \figref{metamorph} in a clearer
context. Since the branch cuts do not play any role now, we relocate
them in such a way that they are not crossed by any of the strings. In
\figref{plato}(a) the network is again shown on the $w$-plane with
its endpoints forming a rectangle whose vertical edges will be
decreased in (b) and (c).  
\onefigure{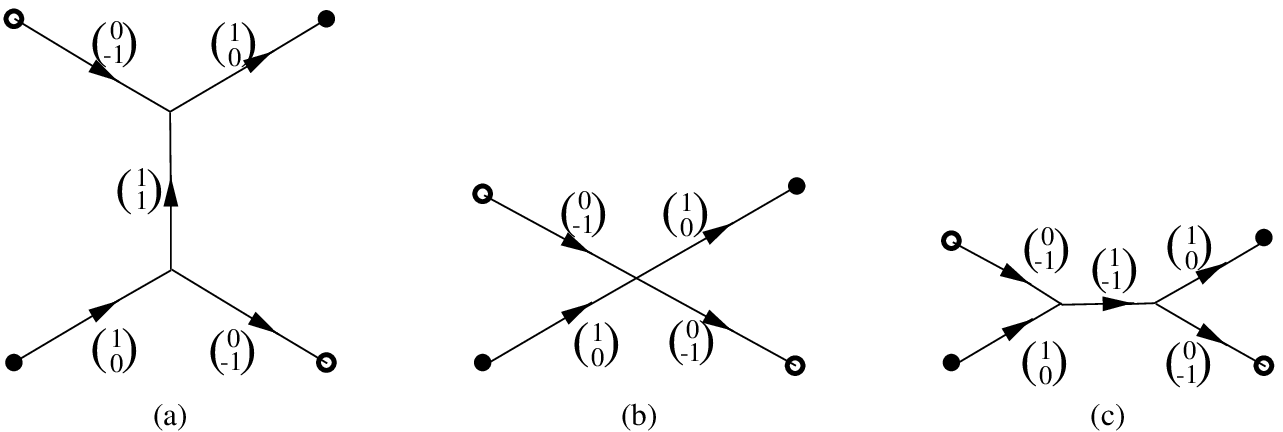}{Channel-transition of a four-pronged network in constant
$\tau=e^{i\frac{\pi}{3}}$ background on the $w$-plane.}

Naively, one could expect a jump in the process of moving the branes
when the {\em square} is reached, since the shortest network seems to become
the one with horizontal string in the middle (this is known as
Plateau's problem). However, we should not forget about the tensions:
although 
\bea
T_{\vect{1}{0}} = T_{\vect{0}{-1}} = T_{\vect{1}{1}},
\eea
the horizontal prong is born with a different value:
\bea
T_{\vect{1}{-1}} = \sqrt{3}T_{\vect{1}{0}},
\eea
and this implies that the angles at the new junction will be
$60^{\circ}$--$150^{\circ}$--$150^{\circ}$, instead of
$120^{\circ}$--$120^{\circ}$--$120^{\circ}$. As a consequence, the
network on \figref{plato}(c) becomes shorter exactly when the
vertical $\vect{1}{1}$-string shrinks to zero size and the transition
happens smoothly through the four-pronged junction of
\figref{plato}(b).

\section{Conclusions}

The conjecture, that string networks appear on equal footing with
conventional open strings in IIB D-brane theories has given rise to a
number of new questions. A great deal of evidence has been accumulated
supporting the necessity of these nonperturbative objects in various
3- and 7-brane configurations, which calls for a consistent
description of the space of these networks. In the present paper we
addressed three issues: the equivalence classes of networks, the
uniqueness of the representative at a point in the moduli space and
transitions between equivalent networks.

In our paper we examined the behavior of $\vect{p}{q}$-strings and
their junction on a nontrivial background created by parallel 7-branes
which also served as endpoints for the network. We first defined what
we mean by the equivalence of graphs and then turned to the question
of which of many equivalent graphs is realized as a BPS network is
realized in a particular background. We analyzed in detail the case of
four equivalent networks in the vicinity of three 7-branes and
concluded that the representative of the equivalence class changes as
we move around in the moduli space but never is more than one realized
in a supersymmetric fashion simultaneously.

In this exercise we learned two important general features about
networks: the calculation of the mass of the network or the slope of a
particular string at a given point is independent of the particular
representative and depends only on the background metric and the
equivalence class. We defined a complex parameter, ${\cal M}$ in terms
of the positions of the 7-branes through which the above dependence
can be simply expressed and found in particular that the BPS mass is
$m = |{\cal M}|$. The determination of the slope of the strings can be
used to prove the uniqueness of the BPS object. We carried out this
argument for the equivalence class of 12 graphs in the vicinity of
four 7-branes and although we did not extend this to the general case,
we suggested a procedure for the construction of a BPS object in an
arbitrary background making use of the invariant charges. 

The most appreciated novelty in these 7-brane configurations is the
appearance of exceptional gauge groups, which require at least eight
7-branes. While having more 7-branes means more complication in
general, transitions on the constant $\tau$ branches can be studied
explicitly. An interesting concrete application was the
channel-transition of the four-pronged network which may be viewed as 
a modification of Plateau's problem of the graphs of least length
between the vertices of a polygon.

\section*{Acknowledgments}

I would like to thank B.~Zwiebach for guidance and suggestions on the
manuscript. I am also happy to acknowledge useful conversations with
O.~DeWolfe and A.~Iqbal.  This work was supported in part by
D.O.E. contract DE-FC02-94ER40818.

\end{document}